\documentclass[preprint,12pt]{elsarticle}
\usepackage{CJK}
\usepackage{amssymb}
\usepackage[pdftex,colorlinks]{hyperref}
\usepackage{multirow}
\usepackage[tbtags]{amsmath}

\biboptions{compress}

\begin{document}

\begin{frontmatter}

\title{A quantum algorithm for approximating the influences of Boolean functions and its applications}

\author{Hong-Wei Li$^{1,2,3,4}$}
\author{Li Yang$^{1,3}$\corref{1}}
\cortext[1]{Corresponding author email: yangli@iie.ac.cn}
\address{1.State Key Laboratory of Information Security, Institute of Information Engineering, Chinese Academy of Sciences, Beijing 100093, China\\
2.School of Mathematics and Statistics, Henan Institute of Education, Zhengzhou,450046,Henan, China\\
3.Data Assurance and Communication Security Research Center, Chinese Academy of Sciences, Beijing 100093, China\\
4.University of Chinese Academy of Sciences, Beijing 100049, China}
\begin{abstract}
We investigate the influences of variables on a Boolean function $f$ based on the quantum Bernstein-Vazirani algorithm. A previous paper (Floess et al. in Math. Struct. in Comp. Science 23: 386, 2013) has proved that if a $n$-variable Boolean function $f(x_1,\ldots,x_n)$ does not depend on an input variable $x_i$, using the Bernstein-Vazirani circuit to $f$ will always obtain an output $y$ that has a $0$ in the $i$th position. We generalize this result and show that after one time running the algorithm, the probability of getting a 1 in each position $i$ is equal to the dependence degree of $f$ on the variable $x_i$, i.e. the influence of $x_i$ on $f$. On this foundation, we give an approximation algorithm to evaluate the influence of any variable on a Boolean function. Next, as an application, we use it to study the Boolean functions with juntas, and construct probabilistic quantum algorithms to learn certain Boolean functions. Compared with the deterministic algorithms given by Floess et al., our probabilistic algorithms are faster.
\end{abstract}


\end{frontmatter}


\section{Introduction}
In \cite{DEM13}, Floess et al. studied the juntas using the quantum Bernstein-Vazirani algorithm, and they proved that using the Bernstein-Vazirani circuit to a Boolean function
$f(x_1,\ldots,x_n)$ that has nothing to do with a variable $x_i$ would always obtain the output $y=(y_1,\ldots,y_n)\in\{0,1\}^n$ with $y_i=0$. In this paper, we will generalize it to the result that the number of ones in each position $i$ of the outputs relates to the influence of $x_i$ on $f$, and our result will contain the one in \cite{DEM13}.

During the last thirty years, there were a lot of researches about the influences of variables on Boolean functions. In \cite{BL89}, the authors transformed randomized algorithms to the processors of flipping the coin. In order that a single processor does not control the global bit, it must have a Boolean function that every variable has a little influence. Hatami \cite{H12} pointed out that the influence of a variable on a Boolean function appears in various contexts such as probability theory, computer science and statistical physics, and Boolean functions with small total influences are in close touch with threshold phenomenon. In \cite{KKL88}, Kahn et al. introduced harmonic analysis methods on Boolean functions for the first time, proved a so called KKL inequality to give a lower bound on total influences. The KKL inequality now is usually used to estimate some bounds \cite{W08,GKKRR08}.

Similar to \cite{DEM13}, our algorithms will be based on the Bernstein-Vazirani algorithm \cite{BV93}, which has the same circuit as the Deutsch-Jozsa algorithm \cite{CEMM98}. And based on cavity QED, some schemes have been proposed to realize Deutsch-Jozsa algorithm \cite{SAG05,HS08}.

In this paper, we begin with some preliminaries. Next, we give a theorem about the influence of a variable of a Boolean function and the gains after running the Bernstein-Vazirani algorithm. Based on this, we propose a quantum approximation algorithm to evaluate the influence, and finally we exploit the above result to the learning of juntas.
\section{Preliminaries}
\subsection{Notations and definitions}
\vspace{2mm}

\noindent{\bf Definition 1}\quad
Let $f(x_1,\ldots,x_n):\{0,1\}^n\rightarrow \{0,1\}$ be a Boolean function, $i\in\{1,2,\ldots,n\}=[n]$, $\alpha^i\in \{0,1\}^n$, and all the coordinates of $\alpha^i$ are $0$ except the $i$th one. For any event $E$, $\mathrm{Pr}(E)$ denotes the probability of $E$ happening. For any set $A$, $|A|$ denotes the cardinality of $A$. \emph{The influence of a variable $x_i$ on the function $f$} is defined as
\begin{equation}\label{eq:a}
I_f(i)=\mathrm{Pr}[f(x\oplus \alpha^i)\neq f(x)]=\frac{|\{x|f(x\oplus \alpha^i)\neq f(x)\}|}{2^n},
\end{equation}
where $\oplus$ denotes bitwise exclusive-or.
\vspace{2mm}

\noindent{\bf Definition 2}\quad For any Boolean function $f$, we define the \emph{Walsh transform} of it by
\begin{equation}\label{eq:b}
S_f(y)=\frac{1}{2^n}\sum_{x\in F^n_2}(-1)^{f(x)+y\cdot x},
\end{equation}
where $y\in \{0,1\}^n$.

\subsection{The Bernstein-Vazirani algorithm\cite{DEM13,BV93,CEMM98,ME11}}
For any Boolean function $f$, define the $U_f$ gate as
\begin{equation}\label{eq:c}
U_f|x\rangle|z\rangle=|x\rangle|z+f(x)\rangle,
\end{equation}
where $x\in \{0,1\}^n$, $z\in \{0,1\}$, and the addition is modulo 2.

Applying $n$ Hadamard gates to $|x\rangle$ obtains
\begin{equation}\label{eq:d}
H^{\otimes n}|x\rangle=\frac{1}{\sqrt{2^n}}\sum_{y\in\{0,1\}^n}
(-1)^{x\cdot y}|y\rangle,
\end{equation}

Now, begin with the initial state $|0\rangle^{\otimes n}|1\rangle$, do the following
\begin{equation}\label{eq:e}
\begin{split}
&\quad|0\rangle^{\otimes n}|1\rangle\\
\underrightarrow{H^{\otimes (n+1)}}
&\quad \frac{1}{\sqrt{2^n}}\sum_{x\in\{0,1\}^n}|x\rangle\cdot \frac{|0\rangle-|1\rangle}{\sqrt{2}}\\
\underrightarrow{U_f}
&\quad \frac{1}{\sqrt{2^n}}\sum_{x\in\{0,1\}^n}(-1)^{f(x)}|x\rangle\cdot \frac{|0\rangle-|1\rangle}{\sqrt{2}}\\
\underrightarrow{H^{\otimes (n+1)}}
&\quad \frac{1}{2^n}\sum_{y\in\{0,1\}^n}\sum_{x\in\{0,1\}^n}(-1)^{f(x)+y\cdot x}
|y\rangle\cdot |1\rangle.
\end{split}
\end{equation}
Discarding the last qubit. If $f(x)=a\cdot x$, $a\in\{0,1\}^n$, we just get $|a\rangle$, and if we measure in the computational basis, we will get $a$.
If $f(x)$ is not a linear function, from \eqref{eq:b} and \eqref{eq:e}, the output is actually
\begin{equation}\label{eq:f}
\sum_{y\in F^n_2}S_f(y)|y\rangle.
\end{equation}
This time if we measure in the computational basis, we will get $y$ with probability $S_f^2(y)$. We will always use $y=(y_1,\ldots y_n)$ to denote the result after running the Bernstein-Vazirani algorithm in this paper.
\subsection{The Hoeffding inequality \cite{WH63}}
If $X_1,X_2,\ldots,X_n$ are independent random variables and $a_i\leqslant X_i\leqslant b_i(i=1,2,\ldots,n),$ then for $t>0$
\begin{equation}\label{eq:af}
\text{Pr}\{|\frac{1}{n}\sum_{i=1}^nX_i-\frac{1}{n}E(\sum_{i=1}^nX_i)|\geqslant t\}
\leqslant 2e^{-2n^2t^2/\sum_{i=1}^n(b_i-a_i)^2},
\end{equation}
where $EX$ is the expected value of the random variable $X$.
\section{The main results about the influences of Boolean functions}
\noindent{\bf Theorem 1}\quad
For any Boolean function $f$,
\begin{equation}\label{eq:g}
I_f(i)=\sum_{y_i=1}S_f^2(y).
\end{equation}

Kahn et al. and O'Donnell have given an expression between the influence of a variable on a Boolean function $f$ and the fourier transform of $f$ in \cite{KKL88} and \cite{RO08}, where the fourier transform of $f$ is
$\widehat{f}(y)=\frac{1}{2^n}\sum_{x\in F^n_2}f(x)(-1)^{y\cdot x}$. We changed it a little so that we could use the Bernstein-Vazirani algorithm to evaluate the influence.
\\
\emph{Proof} (We'll use a method similar to \cite{KKL88} and \cite{LY14} to complete the proof.) First, let
\begin{equation}\label{eq:h}
C_f(\gamma)=\sum_{x\in F^n_2}(-1)^{f(x)+f(x\oplus \gamma)},
\end{equation}
\begin{equation}\label{eq:ah}
S_{(C_f)}(y)=2^{-n}\sum_{\gamma\in F_2^n}C_f(\gamma)(-1)^{\gamma\cdot y}.
\end{equation}
Eq. \eqref{eq:h} is substituted in Eq. \eqref{eq:ah},
\begin{equation}\label{eq:bh}
\begin{split}
S_{(C_f)}(y)
=&2^{-n}\sum_{\gamma\in F_2^n}\sum_{x\in F^n_2}(-1)^{f(x)+f(x\oplus \gamma)}(-1)^{(x\oplus\gamma)\cdot y}(-1)^{x\cdot y}\\
=&2^nS_f^2(y).
\end{split}
\end{equation}
Therefore,
\begin{equation}\label{eq:ch}
\begin{split}
&C_f(\gamma)
=\sum_{y\in F_2^n}S_{(C_f)}(y)(-1)^{\gamma\cdot y}
=2^n\sum_{y\in F_2^n}S_f^2(y)(-1)^{\gamma\cdot y}.
\end{split}
\end{equation}
Thus, we can have
\begin{equation}\label{eq:dh}
C_f(\alpha^i)=2^n(\sum_{y_i=0}S^2_f(y)-\sum_{y_i=1}S^2_f(y)).
\end{equation}
On the other hand,
\begin{equation}\label{eq:i}
\begin{split}
C_f(\alpha^i)=&|\{x \in F^n_2|f(x\oplus \alpha^i)+f(x)=0\}|-|\{x \in F^n_2|f(x\oplus \alpha^i)+f(x)=1\}|\\
=&|V_0|-|V_1|,
\end{split}
\end{equation}
and from \eqref{eq:dh} and \eqref{eq:i}, we have
\begin{equation}\label{eq:j}
\sum_{y_i=0}S^2_f(y)-\sum_{y_i=1}S^2_f(y)=
\frac{|V_0|}{2^n}-\frac{|V_1|}{2^n}.
\end{equation}
In addition, by Parseval's relation, we have
\begin{equation}\label{eq:k}
\sum_{y_i=0}S^2_f(y)+\sum_{y_i=1}S^2_f(y)=1
=\frac{|V_0|}{2^n}+\frac{|V_1|}{2^n}.
\end{equation}
Combining \eqref{eq:j} and \eqref{eq:k}, we can obtain
\begin{equation}\label{eq:l}
\begin{cases}
\sum_{y_i=0}S^2_f(y)=\frac{|V_0|}{2^n},\\
\sum_{y_i=1}S^2_f(y)=\frac{|V_1|}{2^n}.
\end{cases}
\end{equation}
By Definition 1 we have $I_f(i)=\frac{|V_1|}{2^n}$, henceforth
\begin{equation*}
I_f(i)=\sum_{y_i=1}S_f^2(y). 
\end{equation*}

\vspace{2mm}
From Theorem 1, we immediately have the following.

\vspace{2mm}
\noindent{\bf Theorem 2}\quad Using the Bernstein-Vazirani circuit once to a Boolean function $f$, the probability of find the one in a position $i$ (i.e. $y_i=1$) is identical to the influence of $x_i$ on $f$. Specially, if $f$ is independent of $x_i$, we will always find $y_i=0$; if $I_f(i)=1$, we will always find $y_i=1$.
\\
\emph{Proof} \quad According to \eqref{eq:f}, if we measure in the computational basis, we will get $y$ with probability $S_f^2(y)$. So the probability we get $y_i=1$ is
\begin{equation}\label{eq:al}
\text{Pr}(y_i=1)=\sum_{y_i=1}S_f^2(y)
=I_f(i),
\end{equation}
where the second equal comes from \eqref{eq:g}. Meanwhile, the probability we get $y_i=0$ is
\begin{equation}\label{eq:bl}
\text{Pr}(y_i=0)=\sum_{y_i=0}S_f^2(y)
=1-I_f(i).
\end{equation}
Specially, if $f$ is independent of $x_i$, i.e. $I_f(i)=0$, then by \eqref{eq:al} and \eqref{eq:bl}, $\text{Pr}(y_i=1)=0$ and $\text{Pr}(y_i=0)=1$. Consequently, we can not get $y$ with $y_i=1$, we will always find $y$ with $y_i=0$. If $I_f(i)=1$, by \eqref{eq:al} and \eqref{eq:bl}, the probability we get $y_i=1$ is 1, we will always find $y$ with $y_i=1$. 

\vspace{2mm}
\noindent{\bf Remark }\quad \emph{Theorems 1 and 2} conclude \emph{Theorems 3.1 and 3.2} in \cite{DEM13} as a special case. In other words, \emph{Theorems 1 and 2} generalize \emph{Theorems 3.1 and 3.2} in \cite{DEM13} separately.

\section{The quantum algorithm for the influences of Boolean Functions}

\subsection{ The quantum algorithm }

Now, we will give our algorithm. Given an oracle to a Boolean function $f$, our algorithm will run in polynomial times, and output the approximate value of $I_f(i)$ for every $i\in [n]$.

{\bf Algorithm 1}

1.\;Run the Bernstein-Vazirani circuit for the function $f$ $m$ ($m$ is a positive integer independent of $n$) times. Let $y^1, \ldots, y^m$ be the outputs.

2.\;For any fixed $i\in[n]$, count the total number of ones in $y^1_i, \ldots, y^m_i$, and denote it by $l_i$.

3.\;Compute $p_i=p_i(m)=\frac{l_i}{m}$ and output it.

Then by \emph{Theorem 2} \eqref{eq:al},
\begin{equation}
I_f(i)\approx p_i,
\end{equation}
where $\approx$ is an  abbreviation for be approximately equal to.
And the total influence of all variables on $f$ is
\begin{equation}
\sum_{i=1}^nI_f(i)\approx \sum_{i=1}^np_i=\frac{\sum_{i=1}^nl_i}{m}.
\end{equation}
From this we can know some properties about influence, such as which variables have influences more than 0, whether every variable has a little influence or not, whether the total influence is small or not, and so on.

\subsection{ The analysis of Algorithm 1}
What is the error scope of $I_f(i)$ that we just compute through the above method? In other words, what's the distance of $p_i$ and $I_f(i)$?

\vspace{2mm}
\noindent{\bf Theorem 3}\quad $\forall \epsilon> 0$, we have
\begin{equation}\label{eq:n}
\mathrm{Pr}(|I_f(i)-p_i|<\epsilon)>1-2e^{-2m\epsilon^2}.
\end{equation}
\emph{Proof} \quad For any $i\in[n]$, let $Y_i$ be a random variables such that
\begin{equation}
Y_i=
\begin{cases}
1 & y_i=1,\\
0 & y_i=0,
\end{cases}
\end{equation}
where $y_i$ is the $i$th coordinate of the $y$ measured. Then by the Theorem 2,
\begin{equation}
\mathrm{Pr}[Y_i=1]=I_f(i).
\end{equation}
Running the Bernstein-Vazirani  algorithm $m$ times corresponds to $m$ independent identical distributed random variables $Y_i^j$, $j\in[m]$.
By the Hoeffding's inequality,
\begin{equation}\label{eq:m}
\mathrm{Pr}(|I_f(i)-\frac{1}{m}\sum_{j=1}^mY_i^j|<\epsilon)>1-2e^{-2m\epsilon^2}.
\end{equation}
From the second step of the algorithm, we know $\sum_{j=1}^mY_i^j=l_i$, i.e.
\eqref{eq:n} holds. 

\vspace{2mm}
\noindent{\bf Theorem 4}\quad Running the Algorithm 1 can give a list of variables which satisfies

1.\;any variable $x_i$ on the list has the influence $I_f(i)>0$;

2.\;the probability of any variable $x_i$ with $I_f(i)\geqslant \frac{c}{m}$ appearing on the list is at least $1-e^{-c}$ (where $c$ is a constant). \\
\noindent\emph{Proof} \quad  Run the Algorithm 1, If $l_i\geqslant1$, then output $x_i$. This gives a list of $x_i$, we will show the list satisfies the conditions 1 and 2.

By \emph{Theorem 2}, if $I_f(i)=0$, we will find $l_i=0$, $x_i$ can not be on the list. So it must be $I_f(i)>0$ for every $x_i$ on the list.

For $I_f(i)\geqslant \frac{c}{m}$, when we run the Bernstein-Vazirani algorithm once, the probability of getting $y_i=1$ is at least $\frac{c}{m}$, so the probability of getting $y_i=0$ is at most $1-\frac{c}{m}$. Therefore, the probability of always getting $y_i=0$ in step 1 is at most
\begin{equation}\label{eq:am}
(1-\frac{c}{m})^m\leqslant e^{-c},
\end{equation}
so the probability of getting $l_i\geqslant1$ in step 2 is at least
\begin{equation}\label{eq:am}
1-(1-\frac{c}{m})^m\geqslant 1-e^{-c}. 
\end{equation}

\subsection{ Compare with the classical algorithm}
In the classical probabilistic Turing model, if we want to evaluate $I_f(i)$ for arbitrary $i\in[n]$, we should randomly choose a set $A\subset\{0,1\}^n$, and then compute
\begin{equation}
q_i=\frac{|\{x\in A|f(x\oplus \alpha^i)\neq f(x)\}|}{|A|}
\end{equation}
to get a rough estimate, since by \emph{Definition 1}, $I_f(i)\approx q_i$. We define a random variable $Z_i$,
\begin{equation}
Z_i=
\begin{cases}
1 & f(x)\neq f(x\oplus \alpha^i),\\
0 & f(x)= f(x\oplus \alpha^i).
\end{cases}
\end{equation}
Suppose $|A|=m$, then similarly to \emph{Theorem 3}, we can get
\begin{equation}\label{eq:an}
\mathrm{Pr}(|I_f(i)-q_i|<\epsilon)>1-2e^{-2m\epsilon^2}.
\end{equation}
From the above, we can see that for any fixed $i\in[n]$, the classical algorithm can obtain the same accuracy degree as the quantum algorithm. However, running the quantum Algorithm 1 can get the influences of all variables on the function, while the classical algorithm can only get one of it. So our quantum algorithm gains an $O(n)$ times speedup over the classical one.
\section{Applications in some special cases}
Recall that a junta is a Boolean function that only depend on at most $k$ out of $n$ variables, where $k<n$. From \emph{Theorem 2}, the probability of finding the one in the algorithm is only relevant to the influence of the variable, which is entirely unrelated to $k$ and $n$. So we can use the above quantum algorithm to learn juntas. In \cite{DEM13}, Floess et al. examined the quadratic and cubic functions, and gave deterministic quantum algorithms for these functions. They expected to devise probabilistic quantum algorithms. Now, we will complete this work based on the above results of \emph{Theorems 2 and 3}. Before doing this, we need the following lemma.

\noindent{\bf Lemma 1}\quad If $f(x)$ is of the form
\begin{equation}\label{eq:o}
f(x_1,x_2,\ldots x_n)=\prod_{i=1}^rx_i\,\,(r\in[n]),
\end{equation}
then
\begin{equation}\label{eq:p}
I_f(i)=
\begin{cases}
\frac{1}{2^{r-1}} & i\in[r],\\
0 & i\in[n]-[r].
\end{cases}
\end{equation}
\emph{Proof} \quad For $i\in[n]-[r]$, from \eqref{eq:o}, the expression of $f$ does not contain $x_i$ for such $i$, so $f(x)= f(x\oplus \alpha^i)$ for all $x\in\{0,1\}^n$, by \emph{Definition 1}, $I_f(i)=0$.

Obviously $f(x)\neq f(x\oplus \alpha^i)$ if and only if one of them is 0, the other is 1. From \eqref{eq:o}, we have $f(x)=1$ if and only if $x_i=1$ for all $i\in[r]$. So $f(x)\neq f(x\oplus \alpha^i)$ if and only if $i\in[r]$ and $x_j=1$ for $j\in[r]-\{i\}$.
The number of $x\in\{0,1\}^n$ with $x_j=1$, $j\in[r]-\{i\}$ is $2^{n-r+1}$, the total number of $x\in\{0,1\}^n$ is $2^n$, so by \emph{Definition 1} the influence of $x_i$ on the function $f$ is
\begin{equation*}\label{eq:q}
I_f(i)=\frac{2^{n-r+1}}{2^n}=2^{1-r}. 
\end{equation*}

From the proof we can see that the similar conclusion holds for any product
of $r$ variables. Specially, if a variable $x_i$ only appears in linear term, then $I_f(i)=1$. If a variable $x_i$ only appears in quadratic term, then $I_f(i)=\frac{1}{2}$. If a variable $x_i$ only appears in cubic term, then $I_f(i)=\frac{1}{4}$.
\vspace{2mm}

Now we give our probabilistic quantum algorithms.
\subsection{Quadratic functions}
Suppose $f$ is a Boolean function that is composed of linear and quadratic terms and each variable appears in at most one term. Our assignment is to find the variables in linear terms and those in quadratic terms.

{\bf Algorithm 2}

We apply the Bernstein-Vazirani circuit to $f$ $\rho$ ($\rho$ is an integer, and $\rho\geqslant 2$) times, if we always get $1$ in a position $i$, then $x_i$ will be declared to be in linear term. If we get some $1$ and some $0$ in a position $j$, then $x_j$ will be declared to be in quadratic term. If we always get $0$ in a position $k$, then $x_k$ will be declared to be not in the expression of $f$.

Now let us see the success probability of Algorithm 2.

If $x_i$ is in linear term, from \emph{Lemma 1}, $I_f(i)=1$, so by \emph{Theorem 2}, the probability we get $y_i=1$ is 1, we will always find $y_i=1$.

When $x_i$ is in quadratic term, from \emph{Lemma 1}, $I_f(i)=\frac{1}{2}$, running the Bernstein-Vazirani algorithm once, the probability of getting $y_i=1$ ($y_i=0$) is $\frac{1}{2}$, so in Algorithm 2, the probability of getting $y_i=1$ ($y_i=0$) $\rho$ times is
\begin{equation}\label{eq:ar}
(\frac{1}{2})^\rho=\frac{1}{2^\rho}.
\end{equation}

From this we can see that when we declare $x_i$ to be in linear term, $x_i$ may be in quadratic term with a probability $\frac{1}{2^\rho}$, the error probability is exponentially small.

If $x_k$ is declared to be not in the expression of $f$, but in fact $x_k$ will be probably in quadratic term, the probability that this happens is $\frac{1}{2^\rho}$.

If we declare $x_j$ to be in quadratic term, in fact it will be.
But $x_j$ is in fact in quadratic term, the probability that we make a mistake (i.e. we think it is in linear term or not in the expression of $f$) is
\begin{equation}\label{eq:cr}
(\frac{1}{2})^\rho+(\frac{1}{2})^\rho=\frac{1}{2^{\rho-1}}.
\end{equation}

From the above, we can see that all of these error probabilities are exponentially small.
\subsection{Cubic functions}
This time we still suppose $f$ is a Boolean function and each variable appears in at most one term. The difference is that there are cubic terms in the expression of $f$ besides some linear and quadratic terms. Our aim is to determine the variables in linear, quadratic and cubic terms.

{\bf Algorithm 3}

We still just apply the Bernstein-vazirani circuit to $f$ $\lambda$($\lambda$ is a integer independent of $n$ and $\lambda\geqslant 4$) times. If we always get ones in a position $i$, then $x_i$ will be declared to be in linear term. If we find $\mu\lambda$ ones in a position $j$ (where $\mu\in(\frac{1}{2}-\epsilon, \frac{1}{2}+\epsilon)$, $\epsilon$ is a real number and $0<\epsilon<\frac{1}{8}$, we may set $\epsilon=0.1$), then $x_j$ will be declared to be in quadratic term. If we find $\nu\lambda$ ones in a position $l$ (where $\nu\in(\frac{1}{4}-\epsilon, \frac{1}{4}+\epsilon)$, $\epsilon=0.1$), then $x_l$ will be declared to be in cubic term. If we always get $0$ in a position $k$, then $x_k$ will be declared to be not in the expression of $f$.

The analysis of Algorithm 3 will be more complicated than that of Algorithm 2. We give a less precise evaluation. If a variable $x_j \:(x_l)$ is in quadratic (cubic) term, from Lemma 1, $I_f(j)=\frac{1}{2} \:(I_f(l)=\frac{1}{4})$, similarly to \eqref{eq:m}, by Hoeffding's inequality, we have
\begin{equation}\label{eq:am}
\mathrm{Pr}(|\mu-\frac{1}{2}|<\epsilon)>1-2e^{-2\lambda\epsilon^2},
\end{equation}
\begin{equation}\label{eq:bm}
\mathrm{Pr}(|\nu-\frac{1}{4}|<\epsilon)>1-2e^{-2\lambda\epsilon^2}.
\end{equation}
Therefore,  we will draw a conclusion that $x_j \:(x_l)$ is in quadratic (cubic) term with a probability no less than $1-2e^{-2\lambda\epsilon^2}$. The number approximate to 1 exponentially with the increase of $\lambda$. The analysis of linear terms is more like that about Algorithm 2.

\vspace{2mm}
In conclusion, we can use the generalized Bernstein-vazirani quantum for learning some simple functions, such as quadratic, cubic, quartic and maybe higher degree functions. The running time of the algorithm is independent of $n$, just relate to the influence of the variable. Compare with the deterministic quantum algorithm for the cubic function proposed by Floess et al. in \cite{DEM13}, our probabilistic algorithm shows $O(n)$ times speedup.
\section{Conclusions}
We have presented a quantum approximation algorithm to compute the influence of every variable on the Boolean functions. In general, for $n$ variables function, our algorithm is $O(n)$ times faster than the classical one. Moreover, based on this, we give probabilistic quantum algorithms for learning some special functions with simple forms. The running time of our algorithms rely on the forms of the functions, but do not on the total variables of them. So our probabilistic quantum algorithm for cubic functions gives $O(n)$ times speedup over the deterministic quantum algorithm presented in \cite{DEM13}. To this end, we use a similar method to that in \cite{LY14}, but compare with that, the problem investigated here is different from that one, the methods in this paper are less complex than that one. We can also use the Grover-like operator to amplify the amplitude such as \cite{DEM13,ME11}, but this can not bring forth new ideas in the technique. We expect that the methods in this paper will be helpful for some other questions.
\section*{Acknowledgement}

This work was supported by  the National Natural Science Foundation of China under Grant No.61173157.

\end{document}